# Electrical Indicator of Imminent Freezing in Supercooled Water

James D. Brownridge
Department of Physics, Applied Physics, and Astronomy
State University of New York at Binghamton
Binghamton, New York 13902-6016
jdbjdb@binghamton.edu

**Abstract**

Data is presented that demonstrate electrical activity and evidences of dipole alignment in supercooled water and heavy water before and after the onset of freezing. Voltage signals as high as 13 mV have been recorded. In some cases up to 3 seconds before latent heat is released and freezing began. The polarity of the voltage signals is suggestive of molecule dipole alignment prior to freezing.

**Experiment and Discussion**

Supercooled water[1] and supercooled heavy water produce measurable electrical signals just prior to freezing. Data presented in Fig. 1 show the response of two voltmeters and a thermocouple as water began the process of freeze from a supercooled state[2]. Voltage signals as high as 13 mV have been recorded.

These signals were detected with high impedance voltmeters, Keithly model 610C electrometers. The measuring equipment consisted of a glass tube 10 mm in diameter and 25 mm long with ends capped with gold foils that serves as detecting electrodes for the signals. An insulated thermocouple was also inserted into the tube. About 0.5 ml of doubly distilled water was added to the tube with a syringe through a pinhole in one of the gold foils. The tube was suspended horizontally in a thermally insulator copper box that was heated and cooled with thermoelectric modules. The gold foils were connected to ground through a 0.5 MΩ resistor. Voltmeters were connected at the junctions of the electrodes and resistors and the scale selected was from –0.05 V to +0.05 V. The outputs of the thermocouple and voltmeters were connected to a data acquisition system with a resolving time of 100 msec.

When the thermoelectric modules were turned on the copper box was cooled to about –15 °C in less than 5 minutes. The water in the tube cooled at a slower rate, with most samples freezing in 10 to 20 minutes and usually before the water reached –15 °C, in this setup most samples freeze between –2 °C and –12 °C. In Fig. 1 we show water beginning to freeze at ~-2.2 °C. The temperature of the water plus glass tube rose to about -1 °C in 13 seconds. This long lag time (3 sec) between the electrical signal and the slowly rising temperature was most often seen during the 1$^{st}$ or 2$^{nd}$ freezing of a sample. After several freeze and thawing cycles the lag time was reduced to several hundred msec and the rise in temperature of the supercooled water to 0 $^{o}$C occurred in about 100 msec. Similar results was obtained for heavy water. A long lag time was restored on occasion after the water was heated to near boiling.

Using two voltmeters as described above we observed that the polarities of the electrical signals from opposite ends of the tube are always opposite.[3-5] The amplitude of the negative and positive part of each signal changed with time. Often after several freezing and thawing cycles, each end will be 80% to 100% consistent in the polarity of its pulse. This will often be a quasi-stable condition. The meters may be exchanged or one may be removed with no effect on the polarity of the signals. This condition is obtained only when the water in confined in a cylindrical container as described above. The condition of almost consistent polarity is not obtained in containers of other shapes. There is nothing unique about the set-up described here, many variation have been tried and produced similar results. Voltage signals as high as 13 mV have been recorded using other setups and volumes of water larger than a milliliter.

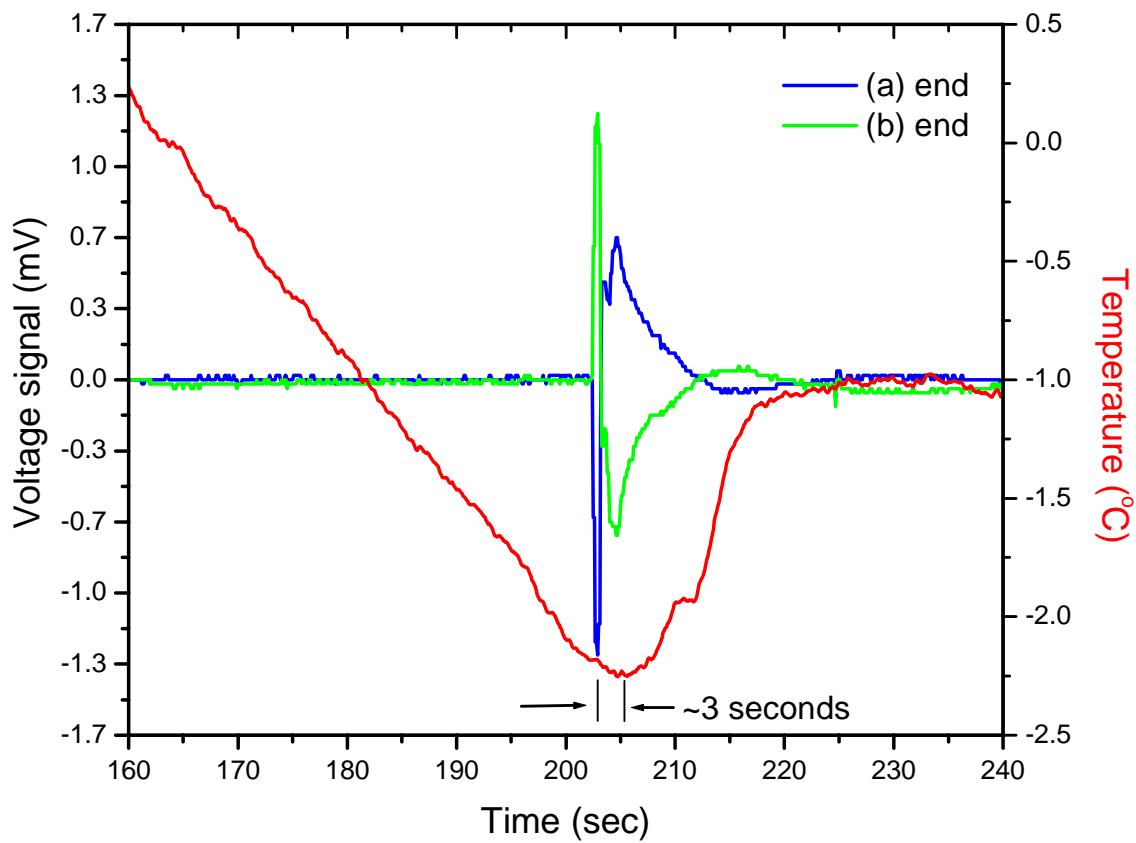

**Figure 1**  Electrical signal from doubly distilled water cooling in environment at -15 °C. The red curve is the water temperature verse time. The blue curve is the voltage signal verse time at one of the glass tube (a). The green curve is the voltage signal verse time at the other end (b). When the water reached minimum temperature of -2.2 °C, latent heat of the water was released and the water temperature began to rise. When the temperature of the water and its container reached –2.2 °C it began to freeze for the first time. Notice that sharp electrical pulses, about +1.2 mV and -1.2 mV are recorded 3 seconds before the water reached -2.2 °C.